\documentclass[12pt]{iopart}
\usepackage{graphicx}
\newcommand\be{\begin{eqnarray}}
\newcommand\ee{\end{eqnarray}}
\newcommand\ba{\begin{array}}
\newcommand\ea{\end{array}}

\begin{document}

\title{Controlling  bi-partite entanglement in multi-qubit systems}
\author{
Martin Plesch${}^{1}$,
Jaroslav Novotn\'{y}${}^{2}$,
Zuzana Dzur\'{a}kov\'{a}${}^{1}$,
Vladim\'{\i}r Bu\v{z}ek${}^{1,3}$}
\address{
${}^{1}$Research Center for Quantum Information, Institute of Physics, Slovak
Academy of Sciences, 845 11 D\'{u}bravsk\'{a} cesta 9, Bratislava,
Slovakia\\
${}^{2}$Department of Physics, FJFI \v{C}VUT,
B\v{r}ehov\'{a} 7,115 19 Praha 1,Czech Republic
\\
${}^{3}$
Faculty of Informatics, Masaryk University, Botanick\'a 68a,
602 00 Brno, Czech Republic
}
\date{13 October 2003}


\begin{abstract}
Bi-partite entanglement in multi-qubit systems cannot be shared
freely. The rules of quantum mechanics impose bounds on how
multi-qubit systems can be correlated. In this paper we utilize a
concept of {\em entangled graphs} with weighted edges in order to
analyze pure quantum states of multi-qubit systems. Here qubits
are represented by vertexes of the graph while the presence of
bi-partite entanglement is represented by an edge between
corresponding vertexes. The weight of each edge is defined to be the
entanglement between the two qubits connected by the edge,
as measured by the concurrence. We prove
that each entangled graph with entanglement bounded by a
specific value of the concurrence can be represented by a {\em
pure} multi-qubit state. In addition we present a logic network
with $O(N^2)$ elementary gates that can be used for
preparation of the weighted entangled graphs of $N$ qubits.
\end{abstract}
\maketitle
\section{Introduction}

Motivated by the seminal paper of Einstein, Podolsky and Rosen
(EPR) \cite{Einstein1935}, Schr\"{o}dinger in his paper entitled
``{\em The present situation in quantum mechanics}''
\cite{Schrodinger1935} has introduced a concept of entanglement.
This new type of purely-quantum mechanical correlation has been
introduced to reflect the fact that (according to Schr\"{o}dinger)
``{\em Maximal knowledge of a total system does not necessarily
include total knowledge of all its parts, not even when these are
fully separated from each other and at the moment are not
influencing each other at all.}'' Quantum correlations have
attracted lot of attention during the history of quantum
mechanics.  Bell \cite{Bell1964} and Clauser {\em et al.}
\cite{Clauser1969} have shown that these correlations violate
inequalities that must be satisfied by any classical local hidden
variable model.

The complex phenomenon of quantum entanglement has been studied
extensively in recent years because it represents an essential
resource for quantum information processing (see, e.g.
reference~\cite{Nielsen2000}). Entanglement between two qubits prepared in
both pure and mixed states is well understood by now. In particular,
necessary and sufficient conditions for the presence of entanglement
in mixed two-qubit states have been derived
\cite{Peres1996,Horodecki1996} and reliable measures of degree of
entanglement have been introduced. Among others, the {\em concurrence}
as introduced by Wootters {\em et al.} \cite{Wootters1997} is a very
useful measure of entanglement since it is rather straightforward to
calculate and is directly related to the entanglement of formation.

Entanglement properties in multi-qubit systems are, on the other hand, still  not completely
revealed.
Firstly, intrinsic multi-partite entanglement is of a totally
different nature then a ``sum'' of bi-partite correlations. Secondly,
unlike classical correlations, bi-partite entanglement cannot be
shared freely among many particles \cite{Coffman2000}. In particular,
Coffman{\em~et~al.}  have derived bounds on bi-partite concurrencies
in three-qubit systems, which are referred to as CKW
(Coffman-Kundu-Wootters) inequalities. Further investigation on
entanglement sharing in multi-qubit systems have been reported in
references~\cite{Wootters2000,OConnor2000,Koashi2000,Dur2001}. In these
papers special states of multi-qubit systems that maximize bi-partite
entanglement between selected pairs of qubits in the system have been
presented. In addition, intrinsic multi-qubit quantum correlations
have been analyzed (see, for instance, references \cite{Miyake2003,Lin2003}).

Controlling the amount of shared bi-partite entanglement in
multi-qubit systems can be used on multi-partite communication
protocols such as quantum secret sharing \cite{Hillery1999} or
specific multi-user teleportation schemes.

The entanglement properties of a multi-qubit system may be represented
mathematically in several ways. D\"{u}r~\cite{Dur2001}, for instance,
has introduced {\it entanglement molecules}: mathematical objects
representing distributions of bi-partite entanglement in a multi-qubit
system.  He has shown that given an entanglement molecule, relevant
\emph{mixed} states with the corresponding entanglement properties can
be found.

An alternative possibility for representing the entanglement relations
of a multi-qubit system is the application of \emph{entangled graphs}.
The entanglement properties of a system with $N$ qubits are represented
by a graph of $N$ vertexes. The vertexes refer to the qubits, while
the edges of the graph represent the presence of entanglement of
the corresponding pairs of qubits.  It was shown in one of our earlier
papers~\cite{Plesch2003} that for every possible graph one can find a
pure state, which would be represented by that graph. The amount of
pairwise entanglement was however not taken into account.

In the present paper we extend the concept of entangled graphs to
describe the amount (degree) of pairwise entanglement in the system as well.
Namely, we assign a weight to each edge of the graph, which is equal
to the amount of the entanglement between the corresponding pair of
qubits. The entanglement is quantified in terms of a concurrence.

For a given state of an $N$ qubit system, one can obviously calculate
pairwise entanglement, thereby constructing the appropriate graph. The
inverse problem, i.e. finding a quantum state with entanglement
properties represented by a given graph, is more difficult.

In Sections 3 and 4 of the paper we will present a complete analysis of existence of
quantum states of multi-qubit systems with entanglement properties
represented by a given particular graph.

For a given graph, many quantum states may be appropriate \emph{per
  se}. The graph itself is not, for instance, sensitive to local
operations on the qubits. On the other hand, there exist graphs for
which no suitable state can be found. The reason behind this is that
bi-partite entanglement cannot be shared freely: e.g. the CKW
inequalities form an obstacle. So, for instance, we cannot have an
entangled graph of three qubits such that each pair is maximally
entangled with the value of concurrence equal to unity.  In spite of
this, a positive statement can be made. We prove in the following,
that if an additional criterion is fulfilled, namely that the weight
of each edge is bounded from above by a certain value, a pure state
corresponding to the given graph can be found. This bound on the
weights depends only on the number of qubits in the system. We also
propose a constructive method, how to find these states.

It is known that an arbitrary quantum state of $N$ qubits can be
prepared using a sequence of single-qubit and two-qubit operations.
These operations can formally be represented as a quantum logic
network. In general one needs to use exponentially many resources
(counted by the number of elementary gates) to prepare a quantum state
of $N$ qubits.

We will show in Section 5, that for preparing a system of qubits in a state
with given entanglement properties resulting from our consideration,
less resources are needed. Namely, a quantum logic network composed of
two- and three-qubit gates enables us to generate the state in
argument. The number of gates building up this network is
\emph{proportional} to the number of entangled qubit pairs in the
system (i.e. the edges of the graph).  In the case of an entangled web
for instance (c.f. reference~\cite{Koashi2000}), when all vertexes of
the graph are connected by edges, the number of gates necessary for a
generation of the state is proportional to $N^2$.

\section{Definitions}
\label{sec2}
\subsection{Concurrence}
\label{Sec2.1}
In this paper we will use concurrence as a measure of bi-partite entanglement.
This has been introduced by Wootters {\it et al.} \cite{Wootters1997} in the
following way: Let us assume a two-qubit system prepared
in a state described by the density operator $\rho $. From this operator one
can evaluate the so-called spin-flipped operator defined as
\begin{equation}
\tilde{\rho}=(\sigma _{y}\otimes \sigma _{y})\rho ^{\ast }(\sigma
_{y}\otimes \sigma _{y}),
\end{equation}
where $\sigma _{y}$ is the Pauli matrix and a star ($^{\ast}$) denotes the
complex conjugation in the computational basis. Now we define the matrix
\begin{equation}
R=\rho \tilde{\rho}\,,  \label{R}
\end{equation}
and label its eigenvalues (which are all non-negative), in decreasing order $\lambda
_{1},\lambda _{2,}\lambda _{3}$ and $\lambda _{4}$. The definition of the concurrence is then
\begin{equation}
C={\rm max}\left\{ 0,\sqrt{\lambda _{1}}-\sqrt{\lambda _{2}}-\sqrt{\lambda
_{3}}-\sqrt{\lambda _{4}}\right\} \,.  \label{C}
\end{equation}
This function also serves as an indicator whether the two-qubit system is
separable (in this case $C=0$), while for $C>0$ it measures the amount of
bipartite entanglement between two qubits with a number between $0$ and $1$.
The larger the value of $C$ the stronger the entanglement between two qubits
is.

\subsection{Coffman-Kundu-Wootters inequalities}
\label{sec2.2}

Coffman {\it et al.}~\cite{Coffman2000} have recently studied a set of
three qubits, and have proved that the sum of the entanglement
measured in terms of the squared concurrence between the qubits $1$
and $2$ and the qubits $1$ and $3$ is less than or equal to the
entanglement between qubit $1$ and the rest of the system, i.e. the
subsystem $23$. Specifically, using the bi-partite
concurrence~(\ref{C}) the state $\varrho_{jk}$ between the qubits $j$
and $k$ we can express the Coffman-Kundu-Wootters (CKW) inequality as
\be
\label{8.1} C_{12}^2+C_{13}^2
\leq C^2_{1,(23)} \, . \ee
Coffmann, Kundu and Wootters have conjectured that
a similar inequality might hold for an arbitrary number $N$ of
qubits prepared in a pure or mixed state. That is, one has
\be
\label{8.2}
\sum_{k=1;k\neq j}^N C^2_{j,k}\leq C^2_{j,\overline{j}}\, ,
\ee
where the sum on the left-hand-side is taken over all qubits
except the qubit $j$, while $C^2_{j,\overline{j}}$ denotes the
concurrence between the qubit $j$ and the rest of the system
(denoted as $\overline{j}$). The maximal value of the concurrence
$C^2_{j,\overline{j}}$ at the right-hand side of equation~\ref{8.2} is equal to
unity.


\subsection{Entangled graphs}
\label{sec2.3}

Let us consider a system of $N$ qubits. As already mentioned, we will
represent the entanglement properties of the system with a weighted
graph with $N$ vertexes. Every qubit is identified with one of the
vertexes, whereas the concurrence between a pair of qubits is
identified with a weighted edge, connecting relevant vertexes. If a
pair of qubits is not entangled at all, there is no edge present in
the graph between the relevant vertexes (thus, the edge with a zero
weight is equivalent to no edge). The graph itself is defined by the number of qubits $N$ and a set of real
numbers $C_{ij}$, giving the concurrencies between relevant pairs of qubits.


\section{Simple examples}
\label{sec3}

The simplest example of a multi-qubit system with interesting
correlation properties was studied in the work of Koashi {\it et al.}
\cite{Koashi2000}. These authors have studied a completely symmetric
state of $N$ qubits such that all $N(N-1)/2$ pairs of qubits in the
system are entangled with the same degree of entanglement.  It has
been shown that a state satisfying this condition is the so-called
 $W$-state
defined as
\begin{equation}
|W\rangle =|N;1\rangle ,  \label{W_stav}
\end{equation}
where $|N;k\rangle $ is a totally symmetric state of $N$ qubits, with $k$
qubits in the state $|1\rangle$ and all the others in the state $|0\rangle
$. The concurrence in this case takes the value
\begin{equation}
C_{\max }=\frac{2}{N},  \label{C_W_max}
\end{equation}
that is maximal under given conditions.

One can easily generalize this example for other completely symmetric configurations
(e.g. for graphs with weights equal on all edges).
As proved by Koashi {\em et al.} \cite{Koashi2000},
if the value of
concurrence is larger than $2/N$ [see equation~(\ref{C_W_max})], then
the desired state does not
exist. If it is smaller than $2/N$ then a pure state corresponding to the
desired entangled web reads
\[
|\Psi \rangle =\sqrt{1-\alpha ^{2}}|N;0\rangle +\alpha |N;1\rangle\, .
\]
The desired value of the concurrence $C$ determines the value of a real
parameter $\alpha$ which reads
\[
\alpha =\sqrt{\frac{CN}{2}}=\sqrt{\frac{C}{C_{\max }}}\, .
\]

A more complicated two-parameter example is the case of a star-shaped
entangled graph (see reference~\cite{Plesch2002}). In this graph a given
qubit is entangled with all the other qubits in the system while no
other qubits are entangled between themselves. In addition, it is
assumed that the strength of the entanglement between the given qubit
and any other qubit is the same (constant). \footnote{ This is a
  special case of o more general graph such that all qubits are
  entangled (kind of an entangled web \cite{Koashi2000}), but one
  qubit (let us denote it as the ``first'' qubit) is entangled with
  the rest of the qubits with the constant concurrence $C_1$, while
  other qubits in the system are mutually entangled as well,
  but the value of the concurrence $C_2$ is different from $C_1$.} In the
reference~\cite{Plesch2002} it has been shown, that asymptotically, in the
limit of large number of qubits (i.e. $N\to\infty$), one is able to
find a state that saturates the CKW inequalities. Thus we are able to
find a state for every
star-shaped graph in the $%
N\rightarrow \infty $ limit \footnote{The upper bound for bipartite
  entanglement given by the CKW inequalities is $C \leq
  \frac{2}{N}$. The upper bound for star-shaped graph is
  $C_{max}=\frac{2}{N} - \delta$, where $\delta \propto
  \frac{1}{N^{2}}$.}.

\section{General solution}
\label{sec4}

As we have mentioned earlier, it has been conjectured, that all
$N$-qubit states have to fulfil the Coffman-Kundu-Wootters (CKW)
inequalities (see equation~(\ref{8.2})) which in the case when the qubit
$j$ is maximally entangled with the rest of the system reads
\begin{equation}
\forall ~j,\quad \sum_{k}C_{kj}^{2} \leq C^{2}_{j,\overline{j}}\leq
1\, . \label{CKW_rovnica}
\end{equation}
Any violation of this inequality means that the corresponding
entangled graph cannot be represented by a quantum-mechanical
state. Under the assumption that all concurrencies $C_{kj}$ in
equation~(\ref{CKW_rovnica}) are mutually equal, i.e. $C\equiv C_{kj}$,
we obtain from the CKW inequality the bound
\[
C\leq \frac{1}{\sqrt{N}}
\]
which is definitely not achievable. To see this we remind ourselves,
that in the case of the entangled web (all qubits are mutually
entangled) the maximal value of the concurrence is given by
equation~(\ref{C_W_max}), which represents a bound that is
much lower than the bound that follows from the CKW inequality.

One may proceed either by deriving tighter CKW-type inequalities
that can be saturated by physical states (graphs).  Alternatively, one
can consider only entangled graphs with specifically bounded weights
on their edges.  In what follows we will study this second option and
will restrict the consideration to those graphs in which the
concurrence on every edge is smaller than a certain value. We will prove that there
exists a nonzero bound on the concurrence such that all graphs with
weighted edges that satisfy this additional condition can be realized
by pure states.

These states are of the form
\begin{equation}
|\Psi \rangle =\alpha |A\rangle
+\sum_{\{i,j\}}\gamma _{ij} |B_{ij}\rangle
\label{Stav}
\end{equation}
where
\be
|B_{ij}\rangle &\equiv&
\left( |
11..0_{i}..0_{j}..1\right\rangle +\left| 00..1_{i}..1_{j}..0\rangle
\right) \, ;
\label{B}
\\
|A\rangle &\equiv&
\left( \left| 00...0\right\rangle +\left| 11...1\right\rangle
\right)\, .
\label{A}
\ee
The real positive coefficients
$\alpha$ and $\gamma_{ij}$ satisfy
a normalization condition
\begin{equation}
2\alpha ^{2}+2\sum_{\{i,j\}}\gamma _{ij}^{2}=1.  \label{Normalizacia}
\end{equation}
The sums in equations~(\ref{Stav}) and (\ref{Normalizacia}) are
 taken through all pairs
$i < j,$ $i,j\in N$ (or, equivalently, the sums can be extended for all pairs $i,j\in N$
with the restriction $\gamma_{ij}=0$ for $j \leq i$). The high (permutational)
 symmetry of the state allows us to
calculate directly the concurrence (for details see Appendix A)
\begin{equation}
C_{ij}=\max \left\{ 2{\left( 2\alpha \gamma
_{ij}-\sum_{k}\gamma _{ki}^{2}-\sum_{k}\gamma
_{kj}^{2}\right) ,0}\right\} ,  \label{Cij}
\end{equation}
which is valid under the condition
\begin{equation}
\alpha \geq 2\gamma _{\max }\sqrt{N-2}\, ,  \label{Podmienka_alfa}
\end{equation}
where
$\gamma _{\max }=\max_{i,j}(\gamma _{ij})$.

Let us note, that the concurrence between every pair of qubits in this
rather complex system is expressed as an analytic function of
input parameters, utilizing just a single condition
(\ref{Podmienka_alfa}).

The set of $\frac{N(N-1)}{2}$ non-linear equations (\ref{Cij}) connect
parameters of the state $\gamma _{ij}$ (the parameter $\alpha $ is
specified by gammas via the normalization condition) with the
concurrencies of different pairs of qubits. This set of equations is
strongly coupled in a sense that in order to calculate one concurrence
one needs to use approximately $2N$ gammas. The task now is to invert
this set of equations, i.e. to find the set of equation defining the
gammas via the set of concurrencies that are given (these
concurrencies do specify the character of the entangled graph).  Not
for every possible choice of concurrencies there exist
parameters $\gamma _{ij}$ satisfying  the normalization condition $%
\Sigma _{i,j}\left| \gamma _{ij}\right| ^{2}< 1$ and the condition
(\ref{Podmienka_alfa}). The reason is that
even though the  concurrencies under consideration have to fulfil the CKW inequalities
these inequalities are just necessary but sufficient condition for the existence of an
 entangled graph
with weighted edges.
Hence, it is also an interesting question, for which set of concurrencies
one can find solutions of the reversed equations (\ref{Cij}).

We have found the solution for the parameters $\gamma_{ij}$
as functions of the concurrencies $C_{ij}$
(weights on the edges of the entangled graph)
that specify the state (\ref{Stav}),
providing all concurrencies are smaller than a certain
maximal value
\begin{equation}
C_{ij}\leq C_{\max },  \label{Podmienka_C}
\end{equation}
where $C_{\max }$ is a given constant.\footnote{The upper bound for
  $C_{\max }$ is obtained from conditions for an iteration procedure
  as defined in Appendix B.}
.\newline

\noindent {\bf Theorem 1}\\
Every entangled graph with weighted edges that is specified by
the set of concurrencies $\{C_{ij}\}$, that fulfil the condition
(\ref{Podmienka_C}), can be represented by a pure state
given by equation~(\ref{Stav})

The complete proof of this Theorem can be found in Appendix B.
Here we just sketch how the
relevant parameters $\gamma _{ij}$ can be obtained via an
iteration algorithm. Let us start from a specific state (\ref{Stav})
corresponding to the situation when
\[
C_{ij}=C_{\max }
\]
for all $i,j$ and then adjusts iteratively the parameters
$\gamma_{ij}$ to fit the
concurrencies. We can  summarize the iteration process as follows:

\begin{itemize}
\item  After each step, all  concurrencies that are evaluated for the
state (\ref{Stav})
are greater than or equal to
the desired  set of concurrencies $C_{ij}$.

\item  After each step, all gammas are smaller than
or equal to their values at the previous step; they do
not change only if for a specific $i,j$ the relevant concurrence is reached.

\item  The iteration limit, when all gammas are zero, leads to zero
concurrencies, too. Therefore, one has to cross the searched state during the
iteration procedure
(for a finite precision this stage can be achieved
after a finite number of iteration steps)
\end{itemize}

The existence of the state itself is proved by showing, that the iteration
process has a proper limit. Also, to ensure the validity of the proposed
process, we made a broad numerical test, with varying number of qubits and
the strength of entanglement. In all tested examples that satisfied the
condition (\ref{Podmienka_C}), a very rapid convergence was observed, when a
precision of about $10^{-6}$ \ of the maximal permitted concurrence was
achieved after nine to twelve steps (changing all gammas at once).

\section{Preparation of entangled graph with weighted edges}
\label{sec5}

In the previous section we have shown that a large class of entangled
graphs with weighted edges can be represented by a pure state~(\ref{Stav}).
It is well known (see e.g. Ref.~\cite{Nielsen2000}) that any state of a
multi-qubit system can be prepared with the help of a
suitable logic network. However, in general
the number of two-qubit gates in
this network increases exponentially with the number of qubits.

In what follows we present a quantum logic network for preparation of
the state (\ref{Stav}), corresponding to a given weighted entangled
graph. This network is very efficient in a sense that it uses only a
quadratic number of three-partite gates with respect to the number of
qubits (every three-qubit gate can be decomposed in at most eight
two-qubit gates).  Three ancilla qubits are needed for the
procedure; these are not entangled with the other ones at the end
of the preparation process.
This keeps the fidelity of the preparation (in the case of error-free
gates) perfect.

\subsection{Definitions}
\label{Siet_def}
Firstly let us introduce  logic gates that will be used
in our network. The first gate is a two-qubit operator, the well-known
controlled NOT ($cNOT$) gate. In this gate the first input qubit serves as a
control. The NOT
operation is applied on the second qubit  when
the control qubit is in
the state $\left| 1\right\rangle $, otherwise  the second qubit does not
change. The operator which implements this gate acts
on the basis vectors of the two qubits under consideration as follows
\begin{eqnarray}
cNOT\left| 0\right\rangle _{i}\left| 0\right\rangle _{j} &=&\left|
0\right\rangle _{i}\left| 0\right\rangle _{j}  \label{Siet_cNOT} \\
cNOT\left| 0\right\rangle _{i}\left| 1\right\rangle _{j} &=&\left|
0\right\rangle _{i}\left| 1\right\rangle _{j}  \nonumber \\
cNOT\left| 1\right\rangle _{i}\left| 0\right\rangle _{j} &=&\left|
1\right\rangle _{i}\left| 1\right\rangle _{j}  \nonumber \\
cNOT\left| 1\right\rangle _{i}\left| 1\right\rangle _{j} &=&\left|
1\right\rangle _{i}\left| 0\right\rangle _{j},  \nonumber
\end{eqnarray}
where $i$ denotes the control and $j$ the target qubit.

The second gate we are going to use is a
three-qubit Toffoli gate $T$ with two control qubits.
In the case that these two control qubits are in the state
$\left| 11\right\rangle $ then  the NOT operation is applied on
the third qubit.
In all other cases the Toffoli gate acts as an identity operator.

The third gate we will use is also a three-qubit gate,
denoted as $R\left( \alpha
\right) $. Here  one qubit will serve as a control.
When this control qubit is in the state $|1\rangle$ then a
specific  ``rotation''
in the two-dimensional subspace of the Hilbert space of the two target qubits will be applied.
This rotation acts on the two target qubits as follows:
\begin{eqnarray}
R\left( \alpha \right) \left| 00\right\rangle &=&\left( 1-\left( \alpha
\right) ^{2}\right) ^{1/2}\left| 00\right\rangle -\alpha \left|
11\right\rangle  \label{Siet_R}\, ; \\
R\left( \alpha \right) \left| 11\right\rangle &=&\alpha \left|
00\right\rangle +\left( 1-\left( \alpha \right) ^{2}\right) ^{1/2}\left|
11\right\rangle  \nonumber\, ; \\
R\left( \alpha \right) \left| 01\right\rangle &=&\left| 01\right\rangle
\nonumber\, ; \\
R\left( \alpha \right) \left| 10\right\rangle &=&\left| 10\right\rangle ,
\nonumber
\end{eqnarray}
where $\alpha $ is a parameter of the rotation.

The last gate that will be used in our network is a three-qubit gate,
which will be denoted by $A$. Again, one qubit serves as a control. The following
transformation is applied to the two remaining qubits when the control qubit is in the state $\left|
  1\right\rangle $:

\begin{eqnarray}
A\left| 00\right\rangle &=&k_{+}\left| 00\right\rangle -k_{+}\left|
01\right\rangle -k_{-}\left| 10\right\rangle +k_{-}\left| 11\right\rangle
\label{Siet_A}\, ; \\
A\left| 01\right\rangle &=&k_{-}\left| 00\right\rangle +k_{+}\left|
01\right\rangle -k_{-}\left| 10\right\rangle +k_{+}\left| 11\right\rangle
\nonumber\, ; \\
A\left| 10\right\rangle &=&k_{+}\left| 00\right\rangle +k_{-}\left|
01\right\rangle +k_{+}\left| 10\right\rangle -k_{-}\left| 11\right\rangle
\nonumber\, ;  \\
A\left| 11\right\rangle &=&-k_{-}\left| 00\right\rangle -k_{-}\left|
01\right\rangle +k_{+}\left| 10\right\rangle +k_{+}\left| 11\right\rangle
\, ,
\nonumber
\end{eqnarray}
where we have used a  short-hand notation $k_{\pm }=\frac{1}{2%
}\sqrt{1\pm \frac{1}{\sqrt{2}}}$.
This operation
will be used in our network  even number of times, so only the effects of
the operation $A^2$ will appear at the end. The operation $A^{2}$ acts
in a simpler and understandable way
\begin{eqnarray}
A^{2}\left| 00\right\rangle &=&-\left| 01\right\rangle  \label{Siet_AA}
\, ;\\
A^{2}\left| 01\right\rangle &=&\left| 11\right\rangle\, ;  \nonumber \\
A^{2}\left| 10\right\rangle &=&\left| 00\right\rangle\, ;  \nonumber \\
A^{2}\left| 11\right\rangle &=&\left| 10\right\rangle\, .  \nonumber
\end{eqnarray}

\subsection{Initial state of the $N$ qubits}
\label{sec5.2}

In order to prepare an entangled graph with $N$ vertexes, i.e.  a
specific $N$-qubit state, we will need three additional ancilla
qubits.  The ancilla is initially prepared in the product state
$|1\rangle|0\rangle|0\rangle$ and is completely factorized from the
other, \emph{graph} qubits. These graph state are initially prepared in the
generalized Greenberger-Horne-Zeilinger (GHZ) state \footnote{To
  generate a GHZ state  one can start with a
  product state of $N$ qubits with the first qubit  in the state $(|1\rangle -
  |0\rangle)/\sqrt{2}$ while all other qubits are in the state$|0\rangle$. Then
  one applies a $cNOT$ gate to every qubit except the first one with
  the control on the first qubit. So, one needs only $N-1$ two-qubit
  gates to prepare the input state.}$(\left| \Psi \right\rangle
_{I}=|1\rangle^{\otimes N} - |0\rangle^{\otimes N})/\sqrt{2}$.  Thus
the input state of the quantum logic network under consideration reads
\begin{equation}
\left| \Psi \right\rangle _{I}\left| 1\right\rangle _{N+1}\left|
0\right\rangle _{N+2}\left| 0\right\rangle _{N+3}=\frac{1}{\sqrt{2}}\left(
\left| 11...1\right\rangle -\left| 00...0\right\rangle \right) \left|
1\right\rangle _{N+1}\left| 0\right\rangle _{N+2}\left| 0\right\rangle
_{N+3}.  \label{Stav1}
\end{equation}

In what follows we will specify  gates in the network
with three indices, where the
first index specifies the control qubit
(or the first two qubits in the case of the Toffoli gate)
and remaining index(es) determine(s) the target qubit(s) of the operation. In
addition, if there will be some action or control applied on the ancillas, we
will denote their relevant indexes as $N+i$, where $i=1,2,3$ is the
position of the ancilla qubit.

\subsection{The network}
The action of the network can be
divided into two main stages.
In the first stage an entangled state of the graph qubits and the ancilla is
created. This state contains state vectors that are essentially the same as
those in the desired state (\ref{Stav}). In the second stage of the
preparation procedure the ancilla becomes factorized from the graph, which
in turn is prepared in the state (\ref{Stav}).

\begin{figure}[tbp]
\centerline{\includegraphics{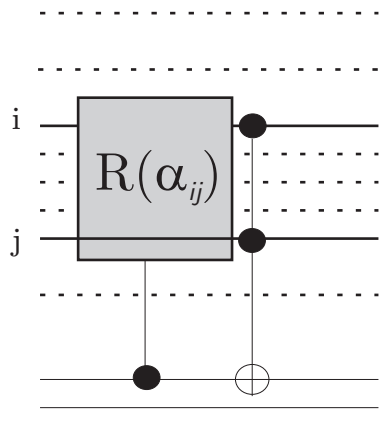}}
 \caption{A schematic description of the logic network corresponding to
the first stage of the preparation procedure.
The rotation $R(\protect\alpha )$
is applied on every pair of the target
qubits (all original qubits, except the ancilla
qubits), with $\protect\alpha $ defined by the parameters
$\protect\gamma _{ij}$ that specify the state corresponding to a given
entangled graph. During this first stage of the preparation procedure
approximately
$N^{2}$ elementary gates are used.}
\label{fig1}
\end{figure}

During
the first stage  of the preparation procedure
we will apply the rotation $R\left( \alpha _{ij}\right) $
to each pair $i\neq j$ from $N$ target qubits with the control on the first
ancilla qubit (see Figure~\ref{fig1}).
After each $R\left( \alpha _{ij}\right)$-gate the Toffoli gate
with the control on $i$
and $j$ qubits
acts on the $first$ ancilla qubit. This
procedure is repeated $
{N \choose 2}$-times,
for all indices $i\neq j$. During each rotation, a fraction
(that is specified by the amplitude $\alpha _{ij}$)
of the state vector $\left( \left|
11...1\right\rangle -\left| 00...0\right\rangle \right) $ is transformed into
the state $\left( \left| 11..0_{i}..0_{j}..1\right\rangle +\left|
00..1_{i}..1_{j}..0\right\rangle \right) $, whereas the already transformed
part of the state $\left( \left| 11..0_{k}..0_{l}..1\right\rangle +\left|
00..1_{k}..1_{l}..0\right\rangle \right) $ is left unchanged.

Thus, after a few steps the state given by equation~(\ref{Stav1}) is transformed
into
\begin{eqnarray}
\left| \Psi \right\rangle &=&\widetilde{\sum_{\{i,j\}}}E_{ij}\left( \left|
11..0_{i}..0_{j}..1\right\rangle +\left| 00..1_{i}..1_{j}..0\right\rangle
\right) \left| 0\right\rangle \left| 0\right\rangle \left| 0\right\rangle \\
&&+D\left( \left| 11...1\right\rangle -\left| 00...0\right\rangle \right)
\left| 1\right\rangle \left| 0\right\rangle \left| 0\right\rangle , \nonumber
\label{Stav2}
\end{eqnarray}
where the tilde indicates that the sum is taken over all  pairs of qubits
 that have already been involved in the transformation.
The corresponding amplitudes
$C_{ij}$ are given by the relation
\begin{equation}
E_{ij}=\frac{\alpha _{ij}}{\sqrt{2}}\left( 1-2\widetilde{\sum_{\{k,l\}}}%
E_{kl}^{2}\right) ^{\frac{1}{2}}  \label{Vaha}
\end{equation}
and $D$ is given by the normalization condition.

We have also to specify the parameters $\alpha _{ij}$ for each rotation.
These parameters are related to amplitudes $\gamma _{ij}$ that
specify the desired state of the entangled graph given by equation~(\ref{Stav}),
which we want to generate.
Comparing the states (\ref{Stav}) and (\ref{Stav2}) we see, that for a
successful generation of the state of the graph
we need  $E_{ij}=\gamma_{ij}$. Using equation~(\ref{Vaha}) we can write
\begin{equation}
\alpha _{ij}=\sqrt{2}\left( 1-2\widetilde{\sum_{\{k,l\}}}\gamma
_{kl}^{2}\right) ^{-\frac{1}{2}}\gamma _{ij}.  \label{Siet_alfa}
\end{equation}

After performing transformations on all pairs of target qubits the resulting
state has the form
\begin{eqnarray}
\left| \Psi \right\rangle =\sum_{\{i,j\}}\gamma _{ij}
|B_{ij}\rangle
\left| 0\right\rangle \left| 0\right\rangle \left| 0\right\rangle
+\alpha
\left| \overline{A}\right\rangle
\left| 1\right\rangle \left| 0\right\rangle \left| 0\right\rangle
.
\label{Stav3}
\end{eqnarray}
where the state vector $|B_{ij}\rangle$ is given by equation~(\ref{B}) and
\be
|\overline{A}\rangle &\equiv&
\left( \left| 11...1\right\rangle -\left| 00...0\right\rangle
\right)\, .
\ee
We see that the component states $|B_{ij}\rangle$ and
$|\overline{A}\rangle$ in equation~(\ref{Stav3}) are essentially the same
as those of the desired entangled graph (see equation~(\ref{Stav})).
Now we will use the first ancilla
qubit the last time before disentangling it from the rest of the system.
We will  apply
the specific controlled rotation on an arbitrary qubit of the graph with
the control being the first qubit of the ancilla. The rotation itself is
described by the operator
$-\sigma _{z}$ (a Pauli matrix). This controlled rotation applied on the state (\ref{Stav3})
performs the transformation
$|\overline{A}\rangle\rightarrow |A\rangle$, while the state
$|B_{ij}\rangle$ remains unchanged.

We see that at this stage the two desired components $\alpha|A\rangle$ and
$\sum\gamma_{ij}|B_{ij}\rangle$  of the graph state
(\ref{Stav}) are generated, but they are entangled with the first ancilla
qubit. The second stage of the preparation procedure is designed so that
the ancilla is disentangled from the graph, while the graph is left in the
state (\ref{Stav}).
To disentangle the first ancilla qubit from the rest of the
system we will use  the other two ancilla qubits.
In order to perform this disentanglement we have to find a network that
will discriminate between two graph states $|A\rangle$ and
$\sum\gamma_{ij}|B_{ij}\rangle$

Let us analyze in more detail the state
$\sum\gamma_{ij}|B_{ij}\rangle$. In this state
two or four neighboring qubits are in different states. This is in contrary to
the state $|A\rangle$,
in which all  qubits are in the same state. The discrimination of the two states can so be performed
by applying the $cNOT$ gate acting always on two neighbouring qubits.
 If the target is
in the state $\left| 0\right\rangle $ then after the action of the $cNOT$ gate
the two qubits that are involved in the action of the gate
are in the same state. On the other hand,  if it is $\left|
1\right\rangle $, the two qubits do differ.
From here it follows that we can use the target qubit as a
control for another gate, which changes its targets if
and only if the two graph qubits do differ.

\begin{figure}[tbp]
\centerline{\includegraphics{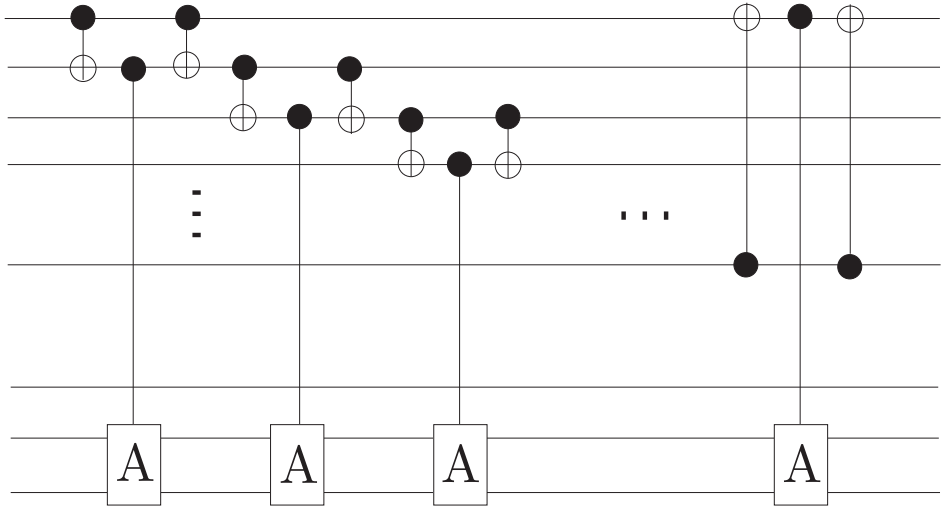}}
 \caption{This part of the network
helps to disentangle the ancilla qubits from the original qubits.
In this case, the rotation $A$ is applied only in the case, when
two neighboring qubits are not equal. Only $N$ elementary gates,
in the order of the magnitude, are used.} \label{fig2}
\end{figure}

Let us utilize for this purpose the gate $A$ which will act on the  last
two ancillas (see Figure~\ref{fig2}).
We apply the $cNOT$ gate on two qubits from $N$ graph
qubits and then we apply the
$A$-gate controlled by the target of the $cNOT$, acting on the last two
ancillas. After that, we apply again the $cNOT$ gate
 on the  same two qubits as before:
This operation will bring all qubits into the original state
(since $cNOT^{2}= I$) and the only effect of this
particular procedure
is a rotation of the state of the last two ancillas. This rotation will
take place
only in the case when the two ``tested'' qubits were in a different state.

Than we repeat the same procedure for each pair of the first $N$ neighboring qubits
of the graph.
After this, the $A$ gate acted either $2$ or $4$ times on  ancilla
qubits, that are entangled with the state $\sum\gamma_{ij} |B_{ij}\rangle$
of the graph. On the other hand
those ancillas that are  entangled with the state $|A\rangle$ are not changed.

The reason for using $A$-gate, the ``square root'' of the operation
(\ref{Siet_AA}), now becomes clear: the $A^{2}$ gate is acting once or
twice and the state $\left|0\right\rangle\left|0\right\rangle$ of the
last two ancillas
is changed either to $%
-\left| 0\right\rangle \left| 1\right\rangle $ or to $\left| 1\right\rangle
\left| 1\right\rangle $.  On the other hand in the case when all target qubits are equal,
$A$ will not act at all and the resulting state of the
last two ancillas will be
unchanged, thus $\left| 0\right\rangle \left| 0\right\rangle $. If the
$cNOT$ gate between the last (control) and the first (target)
ancilla qubit is applied, then
the first ancilla will be changed to the state $\left| 0\right\rangle $ and
it becomes disentangled from the rest of the system. Now all the work is
almost done, the only thing we have to do is to disentangle the two
remaining ancilla qubits. For this we will simply run the procedure for all
neighboring pairs of qubits as described above, but with the gate $A^{-1}$
instead of the gate $A$.
This will change the state of the last two ancilla qubits back
to the original state $\left| 0\right\rangle \left| 0\right\rangle $ and
will finally disentangle the ancilla from the system.
That means that the desired state
$|\Psi \rangle $ of the  $N$ graph qubits is disentangled from the
ancilla and the entangled graph is prepared in the state (\ref{Stav}).

\begin{figure}
\centerline{\includegraphics{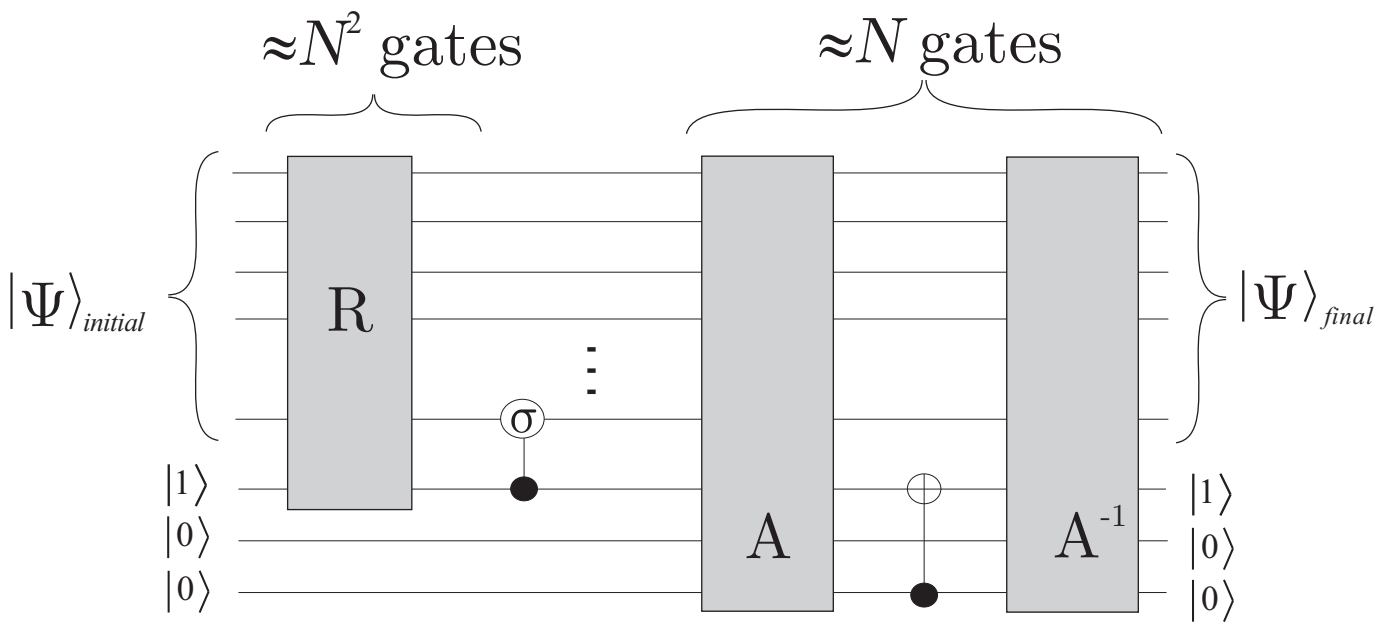}}
\caption{A schematic
description of the entire logic network for the preparation of the
entangled graph in the state (\ref{Stav}). First we use the part
with the rotation $R(\protect\alpha )$, then we correct a sign
with the help of a $cNOT$ gate. Finally, we disentangle the
ancilla qubits with the help of the $A$ and $A^{-1}$ rotations.
The desired state is prepared with the help of (of the order of
magnitude) $N^2$ elementary gates.} \label{fig3}
\end{figure}

Finally, let us summarize the preparation procedure for the
entangled graph given by the state (\ref{Stav}).
As shown in Figure~\ref{fig3},
first we apply the rotations $R$ on all pairs of  qubits of the graph, i.e.
\begin{equation}
G_{1}=\left( c\sigma _{z}\right) _{N+1,N}\prod_{i\neq j}\left(
T_{i,j,N+1}R\left( \alpha _{ij}\right) _{N+1,i,j}\right) ,
\end{equation}
where the subscripts for each operation define the position of qubits, where
operation takes its action. Angles of rotations
$\alpha _{ij}$ are defined  by equation~(\ref{Siet_alfa}) and $c\sigma _{z}$
stands for the controlled sigma gate applied on
the first ancilla as a control
and one of the graph qubits as a target.
At this stage we will use roughly $%
N^{2}$ bipartite gates. The second stage of the preparation corresponds
to disentangling the first
ancilla qubit from the graph qubits
\begin{equation}
G_{2}=cNOT_{N+3,N+1}\prod_{i,i+1}^{N}\left(
cNOT_{i,i+1}A_{i+1,N+2,N+3}cNOT_{i,i+1}\right)\, .
\end{equation}
The last stage of the preparation process
is responsible for disentanglement of the last two ancilla qubits
from the graph qubits, i.e.
\begin{equation}
G_{3}=\prod_{i,i+1}^{N}\left( cNOT_{i,i+1}{A^{-1}}_{i+1,N+2,N+3}cNOT_{i,i+1}%
\right) .
\end{equation}
In the last two equations the indices $i+1$ for the gates are taken implicitly as
modulo $N$. Finally we can represent the action of the whole logic network
as
\begin{equation}
|\Psi \rangle _{F}|1\rangle _{N+1}|0\rangle _{N+2}|0\rangle
_{N+3}=G_{3}.G_{2}.G_{1}.|\Psi \rangle _{I}|1\rangle _{N+1}|0\rangle
_{N+2}|0\rangle _{N+3},
\end{equation}
where $|\Psi\rangle_F$ is the desired state (\ref{Stav})
of the entangled graph with
weighted edges.

\section{Conclusion}

In this paper, we have introduced a concept of entangled graphs with
weighted edges.
Using simple examples we
have shown, that sharing of bipartite entanglement is a complicated
phenomenon and that the
Coffman-Kundu-Wootters inequalities \cite{Coffman2000} are only a
necessary condition for an existence of states with given
entanglement properties.

We have proved that
a whole class of entangled graphs, where the concurrence between an arbitrary
 pair of qubits (vertexes) is weaker than a certain value can be realized by
 a state of $N$ qubits.
Moreover, we have proposed
a logic network for preparation of the states corresponding to this entangled graphs.
The network is composed of  a number of elementary quantum gates that
grows  quadratically  with the number of vertexes (qubits) in the graph.

\ack
This work was supported by the IST-FET-QIPC project QUPRODIS.

\appendix

\section{Concurrence in entangled graphs}

In what follows we will evaluate the concurrence between an arbitrary pair
of qubits of a system in the state (\ref{Stav}), i.e.
\begin{equation}
|\Psi \rangle =\alpha \left( |00...00\rangle +|11...11\rangle
\right)
+\sum_{\{i,j\}}\gamma _{ij}\left( |11\rangle _{ij}|00...00\rangle _{%
\overline{ij}}+|00\rangle _{ij}|11...11\rangle
_{\overline{ij}}\right)\; , \label{StavA}
\end{equation}
where real positive amplitudes $\alpha$ and $\gamma_{ij}$
satisfy the normalization condition
\begin{equation}
2\alpha ^{2}+2\sum_{\{i,j\}}\gamma _{ij}^{2}=1.
\label{NormalizaciaA}
\end{equation}
The sum in equations~(\ref{StavA}) and (\ref{NormalizaciaA}) is taken through all
pairs $i\neq j,$ $i,j\in \hat{N}$, so $\{i,j\}=\{j,i\}$ and thus $\gamma
_{ij}=0$ for $i<j$. The special form of the state (\ref{StavA}) leads to a
rather compact density matrix for an arbitrary two-qubit operator that is
obtained by tracing over the rest of the graph qubits:
\begin{equation}
\rho _{ij}=\left(
\begin{array}{cccc}
A & 0 & 0 & F \\
0 & B & E & 0 \\
0 & E & B & 0 \\
F & 0 & 0 & A
\end{array}
\right)\; ,
\label{Matica}
\end{equation}
where we have used  the notation
\begin{eqnarray}
A &=&\gamma _{ij}^{2}+\alpha ^{2}+\sum_{\{k,l\}}\gamma _{kl}^{2}\, ;
\label{Oznacenie} \\
B &=&\sum_{k}\left( \gamma _{kj}^{2}+\gamma _{ki}^{2}\right)\, ;
\nonumber \\
E &=&2\sum_{k}\gamma _{ki}\gamma _{jk} \, ;
\nonumber \\
F &=&2\alpha \gamma _{ij}\, .  \nonumber
\end{eqnarray}
All sums in equations~(\ref{Oznacenie}) are running through free
parameter(s) $k$ (and $l$), whereas $i$ and $j$  denote a specific
pair of qubits in the graph.
In addition,
the condition $i\neq k\neq l\neq j$ has to be fulfilled.

The convenient form of the matrix (\ref{Matica}) allows us to
calculate square roots of the eigenvalues of the matrix $R$ given by
equation~(\ref{R}):
\begin{eqnarray}
\lambda _{1} &=&A+F  \, ;
\label{Vlastne_hodnoty_2} \\
\lambda _{2} &=&A-F\, ;
\nonumber \\
\lambda _{3} &=&B+E\, ;
\nonumber \\
\lambda _{4} &=&B-E\, .  \nonumber
\end{eqnarray}
Because the coefficients $A,B,E,F$ are positive, the only
candidates for the largest eigenvalue are $\lambda _{1}$ and
$\lambda _{3}$. Let us further define
\begin{equation}
\gamma _{\max }=\max_{i,j}(\gamma _{ij})\, .  \label{gamma_max}
\end{equation}
Using the condition
\begin{equation}
\alpha \geq 2\gamma _{\max }\sqrt{N-2}  \label{Podmienka_alfaA}
\end{equation}
we find
$\lambda _{1}\geq \lambda _{3}$ and the general
expression for the concurrence associated with edges of the entangled graph
prepared in the state (\ref{StavA}) reads
\begin{equation}
C_{ij}=\max \left\{ 2{\left( 2\alpha \gamma
_{ij}-\sum_{\{k,i\}}\gamma _{ki}^{2}-\sum_{\{k,j\}}\gamma
_{kj}^{2}\right) ,0}\right\} .
\end{equation}

\section{Proof of  Theorem 1: Iterative procedure}

In order to prove  Theorem 1 we first label the set of concurrencies that
determine a given entangled graph by
${\bf C}_{ij}$. We will use
a bold {\bf C}  in order to distinguish these concurrencies
from any intermediate
concurrencies, obtained by searching for the state of the entangled graph.

We will start the iteration procedure  with an initial state of
the entangled graph given by equation~(\ref{Stav}). The amplitudes
$\gamma_{ij}$ are specified by the relation
\begin{equation}
\gamma _{ij}^{(0)}\equiv \frac{\lambda }{\sqrt{2+N(N-1)\lambda ^{2}}}\, ,
\end{equation}
that is, the initial state is completely permutationally symmetric.
The parameter $\lambda $ is defined as
\begin{equation}
\lambda =\frac{\sqrt{4(N-2)^{2}+2N(N-1)}-2(N-2))}{N(N-1)}.
\end{equation}
The
corresponding bi-partite concurrencies can be evaluated straightforwardly
and they read:
\be
C_{ij}^{(0)}&=&
C_{\max }=2\left( \alpha ^{(0)}\gamma _{ij}^{(0)}-2(N-2)(\gamma
_{ij}^{(0)})^{2}\right) \nonumber \\
&=&\frac{\sqrt{6N^{2}-18N+16}-2N+4}{N(N-1)}{}\, .
\ee
We remind ourselves that the parameters
$\alpha ^{(0)}$ and $\gamma _{ij}^{(0)}$ are mutually related via
the normalization condition (\ref{Normalizacia}), therefore $\alpha $ is
always implicitly defined by $\gamma _{ij}$.
It is also clear that for the state under consideration
the condition (\ref{Podmienka_alfa}) is fulfilled as well.

Before we describe the iteration procedure itself we introduce the following
notation: we enumerate all pairs of qubits in the entangled graph. All pairs
of qubits (i.e. the edges of the graph) are listed in the set of pairs just
once. At each iteration step one parameter
$\gamma _{kl}$ for a
selected pair of indices $\{k,l\}$ is changed,
whereas all others gammas will stay
unchanged. Let us now suppose, that the $n$-th step of the iteration is done
and both conditions (\ref{Podmienka_C}) and (\ref{Podmienka_alfa})
are still fulfilled.
Moreover $\alpha ^{(n)}$,{}$\gamma _{ij}^{(n)}$ {}are positive.
Hence we find
\begin{equation}
C_{ij}^{(n)}\geq {\bf C}_{ij}\, ; \label{Nerovnost_C}
\end{equation}
\begin{equation}
\alpha ^{(n)}\geq 2\sqrt{N-2}\gamma _{\max }^{(n)}\, ;  \label{Nerovnost_alfa}
\end{equation}
\begin{equation}
0<\alpha ^{(n)}\leq 1\qquad 0\leq \gamma _{ij}^{(n)}<1\, ,
\end{equation}
for all pairs of indices $i$,$j$. The parameter
 $\gamma _{\max }^{(n)}$ is defined
in the same way as in equation~(\ref{gamma_max}), i.e.
\begin{equation}
\gamma _{\max }^{(n)}=\max_{i,j}(\gamma _{ij}^{(n)}).  \label{gamma_max_n}
\end{equation}

In the next iteration step we
take a pair of qubits (i.e., the edge), that follows
after the pair, which was selected in
the previous iteration step $n$.
Let us denote this pair with indices $\{i,j\}$. Then,
in the $(n+1)-st$ iteration step,
we will change the parameters of the state in
a following way:
\begin{equation}
\gamma _{ij}^{(n+1)}=\frac{U^{(n)}-V^{(n)}}{2}\, ;
\label{gamma_n+1}
\end{equation}
\begin{equation}
\alpha ^{(n+1)}=\frac{U^{(n)}+V^{(n)}}{2}\, ,
\end{equation}
where
\begin{equation}
U^{(n)} =\left[(\alpha^{(n)}+\gamma _{ij}^{(n)})^{2}+
\frac{1}{2}({\bf C}_{ij}-C_{ij}^{(n)})\right]^{1/2}\, ;
\nonumber
\end{equation}
\begin{equation}
V^{(n)} =\left[(\alpha^{(n)}-\gamma_{ij}^{(n)})^{2}-
\frac{1}{2}({\bf C}_{ij}-C_{ij}^{(n)})\right]^{1/2}\, .
\nonumber
\end{equation}
All other
gammas remain  unchanged at this iteration step.
Following the conditions (\ref{Nerovnost_C}){%
\ }and (\ref{Nerovnost_alfa}) this iteration step is well defined. Now we
will discuss several important properties of the iteration process:

\begin{itemize}
\item[(1)]  $\alpha ^{(n+1)}$ and $\gamma _{ij}^{(n+1)}$ are solutions of
the equation
\begin{equation}
\alpha ^{(n+1)}\gamma _{ij}^{(n+1)}=\alpha ^{(n)}\gamma _{ij}^{(n)}+\frac{1}{%
4}\left( {\bf C}_{ij}-C_{ij}^{(n)}\right)  \label{gamma_N+1}
\end{equation}
and thus according to equation~(\ref{Cij})
\begin{eqnarray}
C_{ij}^{(n+1)} &=&\max \left\{ 2{\left( 2\alpha ^{(n+1)}\gamma
_{ij}^{(n+1)}-\sum_{\{k,i\}}(\gamma _{ki}^{(n+1)})^{2}-\sum_{\{k,j\}}(\gamma
_{kj}^{(n+1)})^{2}\right) ,0}\right\}  \nonumber \\
&=&\max \left\{ {\bf C}_{ij},0\right\} ={\bf C}_{ij}.
\end{eqnarray}

\item[(2)]  $\alpha ^{(n+1)}$ and $\gamma _{ij}^{(n+1)}$ fulfil
the normalization condition (\ref{Normalizacia}).

\item[(3)]  $\gamma _{ij}^{(n+1)}$ and $\alpha ^{(n+1)}$ are positive and
satisfy the relations
\begin{eqnarray}
0 &\leq &\gamma _{ij}^{(n+1)}<\gamma _{ij}^{(n)}  \label{zmena_gamma}\, ; \\
\alpha ^{(n)} &<&\alpha ^{(n+1)}\leq 1.  \label{zmena_C}
\end{eqnarray}

\item[(4)]  From equations~(\ref{zmena_gamma}) and (\ref{zmena_C}) it
follows that
\begin{equation}
\alpha ^{(n+1)}>\alpha ^{(n)}\geq 2\sqrt{N-2}{}\gamma _{\max }^{(n)}\geq 2%
\sqrt{N-2}{}\gamma _{\max }^{(n+1)}.
\end{equation}
Therefore the condition (\ref{Nerovnost_alfa}) is valid also for the
$(n+1)-st$ iteration step.

\item[(5)]  Let us now show, how will particular concurrencies change in
this single iteration step. For $k,l\neq i,j$ we find
\begin{eqnarray}
C_{kl}^{(n+1)} &=&2\left( 2\alpha ^{(n+1)}\gamma
_{kl}^{(n+1)}-\sum_{\{k,m\}}\left( \gamma _{km}^{(n+1)}\right)
^{2}-\sum_{\{l,m\}}\left( \gamma _{lm}^{(n+1)}\right) ^{2}\right)  \label{1}
\\
&=&2\left( 2\alpha ^{(n+1)}\gamma _{kl}^{(n)}-\sum_{\{k,m\}}\left( \gamma
_{km}^{(n)}\right) ^{2}-\sum_{\{l,m\}}\left( \gamma _{lm}^{(n)}\right)
^{2}\right)  \nonumber \\
&>&2\left( 2\alpha ^{(n)}\gamma _{kl}^{(n)}-\sum_{\{k,m\}}\left( \gamma
_{km}^{(n)}\right) ^{2}-\sum_{\{l,m\}}\left( \gamma _{lm}^{(n)}\right)
^{2}\right)  \nonumber \\
&=&C_{kl}^{(n)}  \nonumber
\end{eqnarray}
and for $k=i$
\begin{eqnarray}
C_{il}^{(n+1)} &=&2\left( 2\alpha ^{(n+1)}\gamma
_{il}^{(n+1)}-\sum_{\{i,m\}}\left( \gamma _{im}^{(n+1)}\right)
^{2}-\sum_{\{l,m\}}\left( \gamma _{lm}^{(n+1)}\right) ^{2}\right)  \label{2}
\\
&=&2\left( 2\alpha ^{(n+1)}\gamma _{il}^{(n)}-\sum_{\{i,m\}}\left( \gamma
_{im}^{(n+1)}\right) ^{2}-\sum_{\{l,m\}}\left( \gamma _{lm}^{(n)}\right)
^{2}\right)  \nonumber \\
&>&2\left( 2\alpha ^{(n)}\gamma _{kl}^{(n)}-\sum_{\{i,m\}}\left( \gamma
_{im}^{(n)}\right) ^{2}-\sum_{\{l,m\}}\left( \gamma _{lm}^{(n)}\right)
^{2}\right)  \nonumber \\
&=&C_{il}^{(n)}.  \nonumber
\end{eqnarray}
The same is valid also for $k=j$.
\end{itemize}

Thus we have shown, that after this iteration step the concurrence for fixed $%
i,j$ (i.e. for the given edge)
will be $C_{ij}^{(n+1)}={\bf C}_{ij}$ and all  other concurrencies of the
entangled graph
will become larger. Thus, the condition for all $i,j$ $%
C_{ij}^{(n+1)}\geq {\bf C}_{ij}$ will be fulfilled. Therefore, the state
defined by equation~(\ref{Stav}) with the
parameters $\gamma _{ij}^{(n+1)}$ can be used
for the next $(n+2)$-nd iteration step.

Therefore, the whole iteration is well defined and we will obtain an
infinite sequence of parameters
$\left\{ \alpha ^{(n)}\right\} _{n=0}^{\infty }$ and $\left\{ \gamma
_{ij}^{(n)}\right\} _{n=0}^{\infty }$ for each pair of indices $i$,$j$
(i.e. for each edge of the entangled graph).
All sequences are monotonous and are bounded, and therefore they
have proper limits. Let us
denote these limits as $\alpha $ and $\gamma _{ij}$
\begin{eqnarray}
\alpha &=&\lim_{n\rightarrow \infty }\alpha ^{(n)}\qquad \Rightarrow \qquad
\alpha \in (0,1\rangle  \label{limita_alfa} \\
\gamma _{ij} &=&\lim_{n\rightarrow \infty }\gamma _{ij}^{(n)}\qquad
\Rightarrow \qquad \gamma _{ij}\in \langle 0,1).  \label{limita_gamma}
\end{eqnarray}

Now we will choose and fix one pair of indices $i$,$j$ and we will show,
that
\begin{equation}
\lim_{n\rightarrow \infty }C_{ij}^{(n)}={\bf C}_{ij}.  \label{limita_C_ij}
\end{equation}
First we define a sequence $\left\{ k(n)\right\} _{n=0}^{\infty }$ in a
following way: $k(1)=p$, where $p$ is a rank of $\{i,j\}$ in the order of
pairs of indices, and $k(n)=p+\frac{nN(N-1)}{2}$. Then
\begin{equation}
C_{ij}^{(k(n))}={\bf C}_{ij}.  \label{vybrana_C_ij}
\end{equation}
The equation (\ref{limita_C_ij}) is equivalent to the definition
\begin{equation}
\left( \forall \,\varepsilon \in {\bf R},\varepsilon >0\right) \left(
\exists n_{0}\in {\bf N}\right) \left( \forall \,n\in {\bf N},n>n_{0}\right)
\left( |C_{ij}^{(n)}-{\bf C}_{ij}|<\varepsilon \right) .
\label{definice_limity}
\end{equation}
Let us
choose and fix the small parameter
$\varepsilon $. Our task is  to find $n_{0}$, that will have
the property (\ref{definice_limity}). Because all sequences $\left\{ \alpha
^{(n)}\right\} _{n=0}^{\infty }$ and $\left\{ \gamma _{kl}^{(n)}\right\}
_{n=0}^{\infty }$ have a proper limit, they are Cauchy
sequences and therefore
\begin{equation}
\left( \forall \,\tau \in {\bf R},\tau >0\right) \left( \exists m_{0}\in
{\bf N}\right) \left( \forall \,n,m\in {\bf N},n,m>m_{0}\right) \left(
\forall \{k,l\}\right)
\nonumber
\\
\left(
\begin{array}{l}
|\alpha ^{(n)}-\alpha ^{(m)}|<\tau \\
|\gamma _{kl}^{(n)}-\gamma _{kl}^{(m)}|<\tau
\end{array}
\right) \, ,  \label{cauchy}
\end{equation}
where
\begin{equation}
\tau =\frac{\varepsilon }{4N(N-1)}\, .  \label{tau}
\end{equation}
For this $\tau $ there exists such $m_{0}$, that the
 property (\ref{cauchy}) is
fulfilled and we can define $n_0$ as
\begin{equation}
n_{0}\equiv k(m_{0})>m_{0}  \label{n_0}
\end{equation}

Further  we will
calculate the difference $|C_{ij}^{(n+1)}-C_{ij}^{(n)}|$
for $n+1>n_{0}$ and $n+1\notin \left\{ k(n)\right\} _{n=0}^{\infty }$.
The last condition means, that
the $(n+1)-st$ iteration step didn't change
$\gamma _{ij}^{(n)}$. From equations~(\ref{1}) and (\ref{2})
we obtain two options for the difference under consideration, either
\begin{equation}
\left| C_{ij}^{(n+1)}-C_{ij}^{(n)}\right| =4\left| \alpha ^{(n+1)}-\alpha
^{(n)}\right| \left| \gamma _{ij}^{(n)}\right| <4\tau <8\tau\, ,
\end{equation}
or
\begin{eqnarray*}
|C_{ij}^{(n+1)}-C_{ij}^{(n)}| &=&\left| 4(\alpha ^{(n+1)}-\alpha
^{(n)})\gamma _{ij}^{(n)}-2(\gamma _{il}^{(n+1)})^{2}+2(\gamma
_{il}^{(n)})^{2}\right| \\
&<&4\left| \alpha ^{(n+1)}-\alpha ^{(n)}\right| \left| \gamma
_{ij}^{(n)}\right| +2\left| \gamma _{il}^{(n+1)}-\gamma _{il}^{(n)}\right|
\left| \gamma _{il}^{(n+1)}+\gamma _{il}^{(n)}\right| \\
&<&8\tau ,
\end{eqnarray*}
where $\gamma _{il}^{(n)}$ is the parameter, which was changed in
the $(n+1)-st$ iteration step.

Finally, we can say for $n>n_{0}$, if $n\in \left\{ k(n)\right\}
_{n=0}^{\infty }$, then $|C_{ij}^{(n)}-{\bf C}_{ij}|=0$. In the opposite
case there exists such $u\in {\bf N}_{0}$, that
\begin{equation}
n\in \langle k(m_{0}+u),k(m_{0}+u+1)\rangle .
\end{equation}
Thus
\begin{eqnarray*}
\left| C_{ij}^{(n)}-{\bf C}_{ij}\right| &=&\left|
(C_{ij}^{(n)}-C_{ij}^{(n-1)})+(C_{ij}^{(n-1)}-C_{ij}^{(n-2)})+\cdots +%
\underbrace{(C_{ij}^{(k(m_{0}+u))}-{\bf C}_{ij})}_{=0}\right| \\
&<&\frac{8N(N-1)\tau }{2}=\varepsilon .
\end{eqnarray*}
But then it must stand
\begin{eqnarray*}
{\bf C}_{ij}=\lim_{n\rightarrow \infty }C_{ij}^{(n)}&=&\lim_{n\rightarrow
\infty }2{\left( 2\alpha ^{(n)}\gamma _{ij}^{(n)}-\sum_{\{k,i\}}(\gamma
_{ki}^{(n)})^{2}-\sum_{\{k,j\}}(\gamma _{kj}^{(n)})^{2}\right) }\\
&=&2{\left(
2\alpha \gamma _{ij}-\sum_{\{k,i\}}\gamma _{ki}^{2}-\sum_{\{k,j\}}\gamma
_{kj}^{2}\right) }.
\end{eqnarray*}
All other conditions remain fulfilled in the limit form as well.
Because this property is
valid for all pairs of indices, we have found the parameters $\gamma _{ij}$,
that define the state (\ref{Stav}) which corresponds to a given entangled
graph.

\end{document}